\documentclass[apj]{emulateapj}

\shorttitle{Pulsations triggered by H burning in low-mass white dwarfs}
\shortauthors{C\'orsico \& Althaus}

\usepackage{natbib}
\begin{document}

\title{Short-period $g$-mode pulsations in low-mass 
white dwarfs triggered by H shell burning}

\author{A. H. C\'orsico$^{1,2}$ \& L. G. Althaus$^{1,2}$}

\affil{$^1$Grupo de Evoluci\'on Estelar y Pulsaciones. Facultad de 
           Ciencias Astron\'omicas y Geof\'{\i}sicas, 
           Universidad Nacional de La Plata, 
           Paseo del Bosque s/n, 
           (1900) La Plata, 
           Argentina\\
       $^2$Instituto de Astrof\'{\i}sica La Plata, 
           IALP (CCT La Plata), 
           CONICET-UNLP\\
}

\email{acorsico@fcaglp.unlp.edu.ar}

\begin{abstract}
The detection of  pulsations in white  dwarfs  with low mass
offers  the possibility  of probing  their
internal structure  through asteroseismology and  place constraints on
the  binary evolutionary  processes involved  in their  formation.  In
this paper we assess the impact of stable H burning on the pulsational
stability  properties of  low-mass  He-core white  dwarf models 
resulting from binary star  evolutionary calculations.  
We  found  that,  apart  from a  dense
spectrum  of  unstable  radial modes and nonradial $g$-  and $p$-modes  
driven  by  the $\kappa$-mechanism  due to the 
partial ionization of H in the stellar envelope, 
some unstable $g$-modes with short pulsation periods are  powered 
also by H  burning via the   $\varepsilon$-mechanism of 
mode driving.     This   is   the    
first   time   that $\varepsilon$-destabilized  modes are found  in models  
representative of cool white dwarf stars. The short periods 
recently detected in the pulsating low-mass white dwarf SDSS 
J111215.82+111745.0 could constitute the first evidence 
of the existence of stable H burning in these stars, in particular in 
the so-called extremely low-mass white dwarfs.
 
\end{abstract}

%_____________________________________________________________________

\section{Introduction}

Low-mass  ($M_{\star}/M_{\odot}  \lesssim  0.45$) white dwarfs (WD) 
are likely the result of intense mass-loss events  at the  
red giant  branch stage of low-mass  stars in  binary
systems before the He flash  onset \citep{review}. 
Since the He flash does not occur, their cores 
must be made of He, at variance with  average mass 
($M_{\star} \sim  0.6 M_{\odot}$) WDs which are though to 
have C/O cores. In  particular, binary evolution is the  
most likely origin
for  the so-called  extremely low-mass (ELM)  WDs, which  have masses
below $\sim  0.18-0.20 M_{\odot}$. According to detailed evolutionary 
computations \citep{2001MNRAS.323..471A,2013A&A...557A..19A}, 
ELM WDs must harbor very thick H envelopes
able to sustain residual H  nuclear burning via  
$pp$-chain, leading to markedly long evolutionary timescales.

Recently, numerous low-mass WDs, including 
ELM  WDs, have been detected through the ELM survey and the SPY and WASP 
surveys \citep[see][]{2009A&A...505..441K,2010ApJ...723.1072B,2012ApJ...744..142B,2011MNRAS.418.1156M,2011ApJ...727....3K,2012ApJ...751..141K}.
The interest in low-mass WDs has been greatly boosted by the recent 
discovery that some of them pulsate \citep{2012ApJ...750L..28H,2013ApJ...765..102H,2013MNRAS.436.3573H}. 
The discovery of pulsating 
low-mass WDs constitutes an unprecedented opportunity 
for probing their interiors and eventually
to test their formation channels by employing the tools of
asteroseismology. A few theoretical pulsational analysis 
of these stars have been 
performed hitherto, which have yielded very interesting results.
In particular, it has been shown 
that $g$-modes in ELM WDs are restricted mainly to the core regions and 
$p$-modes to the envelope, providing the chance to constrain both 
the core and envelope chemical structure of these stars via 
asteroseismology \citep{2010ApJ...718..441S,2012A&A...547A..96C}. 
Also, many unstable $g$- and $p$-modes excited by the 
$\kappa$-mechanism roughly at the right effective
temperatures and the correct range of the periods observed
in pulsating low-mass WDs have been found by \cite{2012A&A...547A..96C}, 
and later confirmed by \cite{2013ApJ...762...57V}.

In this Letter, we perform a new pulsation stability analysis on
the  recently published set of state-of-the-art evolutionary models of
low-mass He-core WDs of \cite{2013A&A...557A..19A}. We focus here on 
the role  of stable H burning on the driving of pulsations through  the 
$\varepsilon$-mechanism. In this excitation
mechanism, the driving   is due  to the  strong sensitivity  of nuclear
burning on  temperature \citep{1989nos..book.....U,1995ARA&A..33...75G}.  
Our computations show that, in addition to the existence of a  
dense spectrum  of  unstable radial,  $g$-  and $p$-modes  driven  by  the
$\kappa$-mechanism  due to the partial ionization of  H,
some unstable short-period $g$-modes of low radial order exist that 
are mainly destabilized by  H  burning via  the
$\varepsilon$-mechanism.  Recently, \cite{2014arXiv1405.4568M}
have reported the existence of low-order $g$-modes destabilized by 
the $\varepsilon$-mechanism in hot H-rich pre-WD models.
The results of the present paper constitute the first theoretical 
evidence of pulsation modes excited by the $\varepsilon$-mechanism in 
cool WD stars. 

\section{Evolutionary models and numerical tools}  
\label{evolutionary}  

\begin{figure} 
\includegraphics[clip,width=1.05\columnwidth]{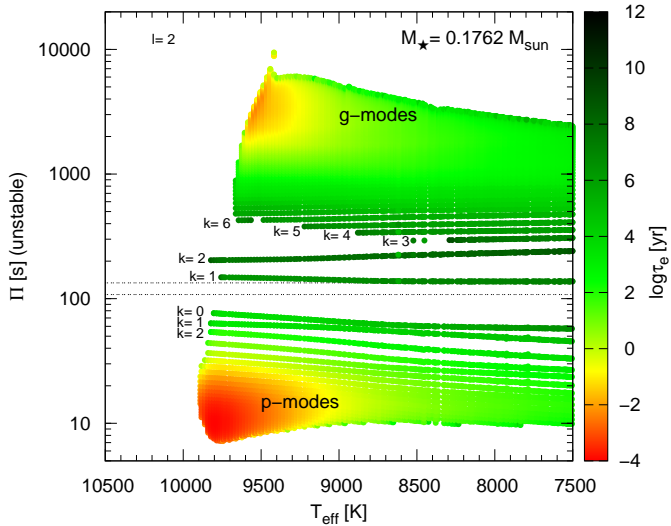} 
\caption{Unstable $\ell= 2$ mode periods ($\Pi$) in terms 
of the effective temperature corresponding to a ELM WD
model sequence with $M_{\star}= 0.1762 M_{\odot}$. Color coding indicates 
the value of the logarithm of the $e$-folding time ($\tau_e$) of each 
unstable mode (right scale).  Horizontal dashed lines correspond 
to the short periods observed in
the pulsating ELM WD star SDSS J111215.82+111745.0, 
at $\Pi \sim 108$ s and $\Pi \sim 134$ s.}
\label{per-teff-te} 
\end{figure} 

Realistic configurations for the low-mass He-core WD
models employed in this work were obtained by \cite{2013A&A...557A..19A}
with the {\tt LPCODE} evolutionary code by mimicking the binary 
evolution of progenitor stars. Specifically, the models were 
derived by computing the non-conservative evolution of a 
binary system consisting of an initially $1 M_{\odot}$ ZAMS star 
and a $1.4 M_{\odot}$ neutron star for various initial orbital periods. 
Details about this procedure can be found in that paper.
Since H shell burning is the main source of star luminosity
during most of the evolution of ELM WDs, the computation
of realistic initial WD structures is a fundamental issue, in
particular concerning the correct assessment of the H envelope
mass left by progenitor evolution. We analyzed six sequences with 
stellar masses of $M_{\star}/M_{\odot}= 0.1554, 0.1650, 0.1762, 0.1806, 0.2707$ 
and $0.4352$. The  pulsation  computations were performed with the help 
of the the linear, radial and nonradial, nonadiabatic versions 
of the {\tt LP-PUL} pulsation code described \cite{2006A&A...458..259C}
 \citep[see also][]{2009ApJ...701.1008C}. 
We have considered $\ell= 1$ and $\ell= 2$ modes. Our computations
ignore  the perturbation of the convective flux; that is, we assume 
the ``frozen-in convection'' approximation.

\section{Nonadiabatic results}
\label{epsilon}

In Fig.  \ref{per-teff-te} we depict
the instability domain of $\ell= 2$ periods  in terms of
the effective temperature for the ELM WD model sequence  
with $M_{\star}= 0.1762 M_{\odot}$. The palette of colors (right scale) 
indicates the value of the logarithm of the $e$-folding time (in yrs)
of each unstable mode, defined as $\tau_{\rm e}= 1/|\Im(\sigma)|$,
where $\Im(\sigma)$  is the imaginary part of the  complex 
eigenfrequency $\sigma$. Many unstable high-order pulsation modes
exist, that are clearly grouped in two separated regions, one of them 
characterized by long periods and associated to $g$-modes, 
and the other one characterized by short periods
and corresponding to $p$-modes. Unstable radial modes 
(not shown in the figure) are also found. Most of these modes are 
destabilized by the $\kappa$-mechanism acting at the surface 
H partial ionization zone. The strongest excitation 
(that is, the smallest $e$-folding times, red zones) is found for 
high-order $g$- and $p$-modes,
with periods in the ranges $[2000-6000]$ s and $[10 - 30]$ s, 
respectively, and effective temperatures 
near the hot boundary of the instability 
islands ($T_{\rm eff} \sim 9700$ K). Similar results,
although with longer unstable $g$-mode periods $(3000-10\,000)$ s, are 
obtained for $\ell= 1$ (not shown). At lower effective
temperatures, these unstable modes become less excited, as reflected
by the higher values of the $e$-folding times ($100-10000$ yrs).
On the other hand, low-order $g$- and $p$-modes and even the $f$-mode  
are also driven, 
although they take much longer to become unstable, as it is reflected
by the dark green tone in the figure ($\tau_e \sim 10^6-10^9$ yrs).
However, since the evolution of the ELM WDs  is so slow, 
these modes would still have enough time as to get excited and reach 
observable amplitudes. This  is confirmed by examining Table
\ref{table1}, in which we show the time $\Delta t$  that the
models take to cool from $T_{\rm eff} \sim  10\, 000$ K to  $\sim
8000$ K, and the maximum $e$-folding times of the unstable short period 
$g$-modes for each stellar mass considered in this work. 
Note that, in all the  cases,
the $e$-folding times are substantially shorter than  the time that
models spent evolving in the regimen of interest. In particular, for
$M_{\star}= 0.1762 M_{\odot}$ the maximum $e$-folding times are a factor $5-100$
shorter than the evolutionary timescale  for the $g$-modes with  $k=
1, \cdots, 5$. 

\begin{table*}
\centering
\caption{The stellar  mass, the mass  of H, the evolutionary 
timescale, the radial order and 
harmonic degree, the $T_{\rm eff}$-range of instability, 
the average period, 
and the maximum $e$-folding time of  
unstable short-period $\ell= 1,2$ $g$-modes destabilized through the
$\epsilon$-mechanism.}
\begin{tabular}{ccccccr}
\hline
\hline
\noalign{\smallskip}
 $M_{\star}$  & $M_{\rm H}/M_{\star}$ &  $\Delta t$   &  $k\ (\ell)$ & $T_{\rm eff}$ & $\langle \Pi\rangle$  &   $\tau_e^{\rm max}$ \\
\noalign{\smallskip}
$[M_{\odot}]$ & $[10^{-3}]$ & $[10^9 {\rm yr}]$ &     &  $[$K$]$     &  $[$s$]$ &   $[10^9 {\rm yr}]$ \\
\noalign{\smallskip}
\hline
\noalign{\smallskip}
0.1554       & 25.4  & 3.13            &  2 (1)  & $\lesssim 8500$ & $350$ & 0.07 \\
             &       &                 &  3 (1)  & $9000-8300$     & $470$ & 0.97  \\ 
             &       &                 &  2 (2)  & $\lesssim 8100$ & $227$ & 0.12  \\  
             &       &                 &  3 (2)  & $8600-8360$     & $291$ & 0.2   \\  
             &       &                 &  4 (2)  & $9000-8800$     & $355$ & 0.33  \\  
\hline
\noalign{\smallskip}
0.1650       & 18.7  & 5.53            &  1 (1) & $\lesssim 8200$ & $250$ & 0.07  \\
             &       &                 &  2 (1) & $\lesssim 9500$ & $340$ & 0.17  \\ 
             &       &                 &  3 (1) & $9500-9020$     & $450$ & 1.3   \\ 
             &       &                 &  4 (1) & $\lesssim 7800$ & $580$ & 0.8   \\ 
             &       &                 &  2 (2) & $\lesssim 8950$ & $214$ & 0.5   \\ 
             &       &                 &  3 (2) & $9400-9300$     & $277$ & 0.5   \\ 
\hline
\noalign{\smallskip}
0.1762       & 14.5 & 7.56             &  1 (1) & $\lesssim 9\,100$ & $247$ &   1.4   \\
             &      &                  &  2 (1) & $\lesssim 10\,000$& $320$ &   0.2   \\
             &      &                  &  3 (1) & $\lesssim 8\,700$ & $470$ &   0.7   \\
             &      &                  &  4 (1) & $8900-8700$       & $550$ &   0.09  \\
             &      &                  &  5 (1) & $9200-9150$       & $620$ &   0.06  \\
             &      &                  &  1 (2) & $\lesssim 8300$   & $140$ &   0.02 \\
             &      &                  &  2 (2) & $\lesssim 9900$   & $220$ &   0.4   \\
             &      &                  &  3 (2) & $8300-7700$       & $297$ &   0.25  \\
\hline
\noalign{\smallskip}
0.1806       & 3.68 & 0.34             &  1 (1) & $\lesssim 10500$   & $270$ &  0.05  \\
             &      &                  &  2 (1) & $10200-9700$       & $355$ &  0.02  \\  
             &      &                  &  1 (2) & $\lesssim 10\,500$ & $178$ &  0.4   \\  
\hline
\noalign{\smallskip}
0.2707       & 1.09 & 0.33             &  $\cdots$  &   $\cdots$     & $\cdots$ &   $\cdots$   \\
\hline
\noalign{\smallskip}
0.4352       & 0.63 & 0.91             & 1  (1) & $\lesssim 9950$    &  $137$ &  0.12  \\
             &      &                  & 1  (2) & $\lesssim 10\,000$ &  $80$  &  0.15  \\
\noalign{\smallskip}
\hline
\hline
\end{tabular}
\label{table1}
\end{table*}

\begin{figure} 
\includegraphics[clip,width=1.05\columnwidth]{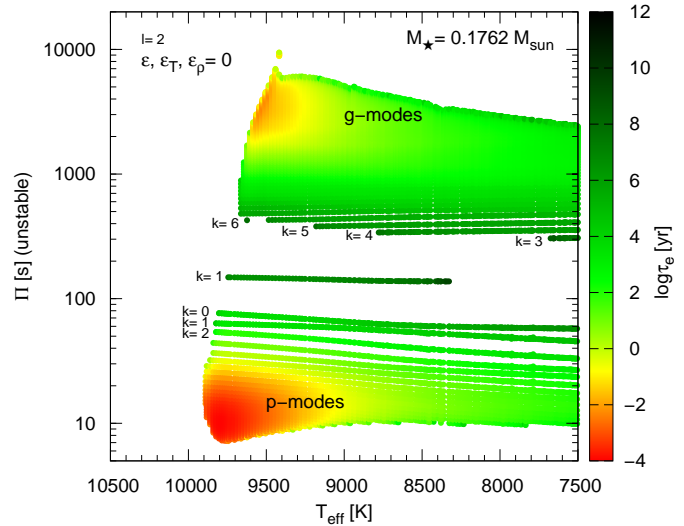} 
\caption{Same as Fig. \ref{per-teff-te}, but for the 
case in which the effects of
the $\varepsilon$-mechanism are suppressed in the 
stability calculations.}
\label{per-teff-te-no-epsilon} 
\end{figure}

At this stage, we can wonder what is the role (if any) of stable H burning 
in the destabilization of the modes shown in Fig. \ref{per-teff-te}. 
To answer this query, we have repeated the stability computations, 
but this time by  consistently suppressing the action 
of this destabilizing agent, that is, 
by forcing the nuclear energy production rate, $\varepsilon$, 
and their logarithmic derivatives 
$\varepsilon_{\rm T}= \left(\frac{\partial \ln 
\varepsilon}{\partial \ln T}\right)_{\rho}$ 
and $\varepsilon_{\rm T}= \left(\frac{\partial \ln 
\varepsilon}{\partial \ln \rho}\right)_{\rm T}$ to be 
zero in the pulsation equations. The results are shown in Fig. 
\ref{per-teff-te-no-epsilon}. Interestingly enough, 
the $k= 2$ $g$-mode
becomes stable and do not appear in this plot. 
Something similar happens with the modes $k = 1$ y $k= 3$
in certain ranges of $T_{\rm eff}$. We can conclude 
that these modes are excited 
(at least in part) by the $\varepsilon$-mechanism through the
H-burning shell. 

\begin{figure} 
\includegraphics[clip,width=1.0\columnwidth]{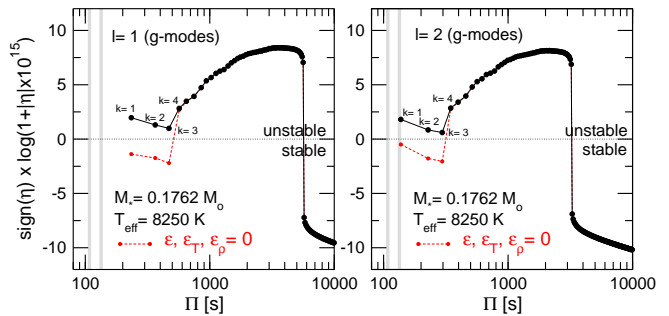} 
\caption{The normalized growth rates 
$\eta$  (large black dots connected with continuous lines) for
$g$-modes in  terms  of  the pulsation  periods for  a $0.1762
M_{\odot}$ ELM WD template model   at $T_{\rm eff}\sim 8\,250$ K.  The large
numerical range spanned by $\eta$ is appropriately  scaled for  a
better graphical  representation \citep[see][]{1997A&A...320..811G}.   
Small red dots  connected with
dashed lines correspond to the case where the $\varepsilon$-mechanism
is explicitly suppressed in the  stability calculations. 
Vertical grey lines correspond to the short periods observed in
the pulsating ELM WD star SDSS J111215.82+111745.0, 
at $\Pi \sim 108$ s and $\Pi \sim 134$ s.}
\label{eta} 
\end{figure}

In what follows, we focus our 
discussion on a template model at $T_{\rm eff} = 8\,250$ K. Fig. \ref{eta}
displays  the normalized  $\ell= 1$  and $\ell= 2$ growth  rates 
$\eta=  -\Re(\sigma)/\Im(\sigma)$ ($\Re(\sigma)$ being the real
part of the  complex eigenfrequency $\sigma$) in terms
of the pulsation periods for this model. 
$\eta > 0$ ($\eta < 0$) implies unstable (stable) modes.  
Large black (small red) dots 
connected with continuous (dashed) lines correspond
to the case where the $\varepsilon$-mechanism is explicitly considered
(suppressed) in the nonadiabatic calculations.  For $\ell= 1$,
the range of periods of unstable $g$-modes is $234-5628$ s
($1 \leq k \leq 57$), while for $\ell= 2$
the periods of unstable modes are in the range  $137-3217$ s 
($1 \leq k \leq 56$). In absence of the destabilizing effect of the 
$\varepsilon$-mechanism, the unstable modes with $k= 1, 2$ and $3$ 
and for both values of the harmonic degree ($\ell= 1$ and $2$) 
turn out to be stable, 
while the remainder modes of the pulsation spectrum ($k \geq 4$) 
remain unchanged. Clearly, the destabilizing effect of H burning is crucial
for the modes with $k= 1,2$ and $3$ to be unstable. Interestingly enough, the 
period of the $\ell= 2$, $k= 1$ mode ($\sim 137$ s) is very 
close to one of the short periods observed in the pulsating 
ELM WD SDSS J111215.82+111745.0 \citep{2013ApJ...765..102H}, 
$\Pi \sim 134$ s (see the rightmost vertical grey line 
in Fig. \ref{eta}). While the star is hotter than our models with the 
$\ell = 2$, $k=1$ $\varepsilon$-destabilized  mode 
($\sim 9590$ K vs $\sim 8300$ K), 
it should be kept in mind that the spectroscopic estimate of 
the effective temperature for this star (like for other ELM WDs) 
can suffer from rather large uncertainties.

\begin{figure} 
\includegraphics[clip,width=1.0\columnwidth]{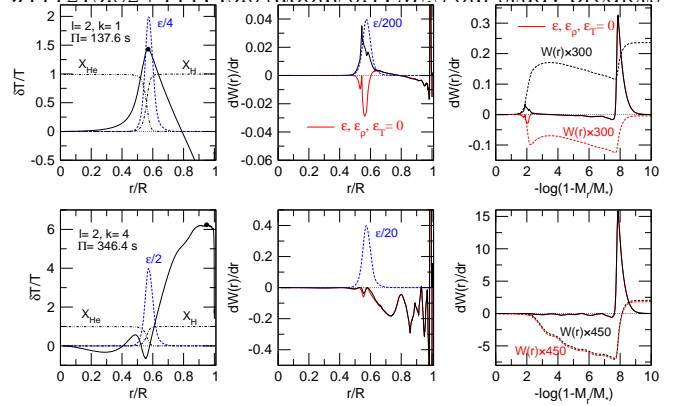} 
\caption{Left panels depict the Lagrangian perturbation of temperature 
($\delta T/T$) along with the scaled nuclear generation rate 
($\varepsilon$) and the H and He chemical abundances 
($X_{\rm H}, X_{\rm He}$) in terms 
of the normalized stellar radius for $\ell= 2$ $g$-modes 
with $k= 1$ (upper panel) and $k= 4$ (lower panel) corresponding to
our template model.
The black dots mark the location of the outermost maximum  of 
$\delta T/T$. Middle panels show the corresponding differential work 
functions ($dW(r)/dr$) for the case in which the $\varepsilon$-mechanism is 
allowed to operate (black curves) and when it is suppressed 
(red curves). Finally, right panels display the same than middle panels, 
but in terms of the mass fraction coordinate. In addition, 
the running work integrals ($W(r)$) are shown.}
\label{dwdr} 
\end{figure}

 In the left panels of Fig. \ref{dwdr} we show the
Lagrangian perturbation of temperature, $\delta T/T$, for 
the $\ell= 2$ modes with $k= 1$ and $k= 4$  
of the template model. 
The peak of the (scaled) nuclear generation rate 
$\varepsilon$  at $r/R \sim 0.57$  marks 
the location of the H-burning shell at the He/H chemical interface.  
We emphasize the position of the outermost maximum of $\delta T/T$
with a black dot. The $\varepsilon$-mechanism behaves
as an efficient filter of modes that provides substantial driving 
only to those $g$-modes that have their largest maximum of 
$\delta T/T$  in the narrow region  of the burning  shell 
\citep{1986ApJ...306L..41K}.  For  our template model, this 
condition is met by the modes with  $k= 1, 2$ and  $3$.   
As $k$ increases, the final
extremum in $\delta T/T$ moves outwards the burning shell. This is the 
case of the mode with $k= 4$, which is not destabilized 
by the H-burning shell at all. The regions of the model that 
contribute to driving and damping for each of the selected modes 
can be drawn with the help of the 
differential work functions, $dW(r)/dr$, which are depicted in the 
middle panels of Fig. \ref{dwdr}. $dW(r)/dr > 0$ implies driving, 
and $dW(r)/dr < 0$ correspond to damping. If the 
$\varepsilon$-mechanism is allowed to operate, there is substantial 
driving for the $k= 1$ mode  
at the location of the 
H-burning shell, and the same happens for the $k= 2$ and $3$ 
modes (not shown). At variance with this, the mode with $k= 4$ 
experiences some damping at that regions. When we set 
$\varepsilon$ and its derivatives to zero 
(red curves), strong damping takes place at that regions for the 
modes with $k= 1, 2$ and $3$, although the situation for the 
mode with $k= 4$ does not change. In the right panels of 
Fig. \ref{dwdr}  we show 
$dW(r)/dr$ and the running work integral, $W(r)$,  
in terms of the coordinate $-\log(1-M_r/M_{\star})$
which strongly amplifies the outer regions of the model. This allow
us to investigate what happens in the outer part of the star, 
where the $\kappa$-mechanism operates due to partial ionization of H.
Notably, there exists strong driving for the two depicted modes due to the 
$\kappa$-mechanism at that regions (something that is barely 
distinguishable in the middle panels because this driving takes place 
at $r/R \sim 1$). Regarding the running work integral, we note that for 
the mode with $k= 1$, $W(r)$ suddenly increases 
at the driving region where the H-burning shell is located,
and also at the outer driving zone associated to the 
$\kappa$-mechanism. The combined effect of both driving regions overcomes
the radiative damping from other parts of the star, and as a result, 
this mode is globally unstable, as proven by the fact that 
$W > 0$ at the stellar surface. The same holds for modes 
with $k= 2$ and $3$. In the case of the mode with $k= 4$, 
the damping effects at the  H-burning shell and  regions 
up to $-\log(1-M_r/M_{\star}) \sim 7$ are important but  not 
enough as to overcome the strong driving at the outer regions, 
and so, it is a unstable mode ($W(1) > 0$). 
If we inhibit the effects of the H burning
(red dashed lines), the mode with $k= 1$ becomes stable, and 
the same occurs for the modes with $k= 2$ and $3$.
We can conclude that, for this specific model, the modes with $k= 1, 2$ and
$3$ are globally unstable thanks to the destabilizing effect of the 
H-burning shell through the $\varepsilon$-mechanism.

\section{Summary and conclusions}

In this work, we have shown for the first time that low-order
short-period $g$-modes are destabilized  through the
$\varepsilon$-mechanism operating at the H-burning shell of cool
low-mass He-core WD models. Note that the $\varepsilon$-mechanism
is responsible for the fact that these modes become pulsationally unstable at
early stages in which the $\kappa$-mechanism is not still efficient
to   drive pulsations at all (see Fig. \ref{per-teff-te} in the
particular case of the sequence with $M_{\star}= 0.1762
M_{\odot}$).  The $e$-folding times of the
$\varepsilon$-destabilized modes are by far shorter than  the
evolutionary timescale (Table \ref{table1}), which  means that they
would have enough time as to get excited  and reach observable
amplitudes.  We find that in low-mass WDs the range of periods
destabilized  by the $\varepsilon$-mechanism is $80 \lesssim \Pi
\lesssim 600$ s; see Table \ref{table1}. In this connection, the
pulsating ELM WD, SDSS J111215.82+111745.0 \citep{2013ApJ...765..102H}, 
that exhibits  two short periods, at
$\Pi \sim 108$ s  and $\Pi \sim 134$ s,  constitutes an
observational counterpart of these theoretically predicted  unstable
modes. Specifically, at the spectroscopic mass inferred for SDSS
J111215.82+111745.0 (about $0.17 M_{\odot}$) our study predicts that
the observed short periods,  in particular that of 134 s, is
reproduced by the period of the excited mode with  $\ell =
2$, $k=1$ (see Table \ref{table1} and Fig. \ref{eta}), 
though our models with this mode destabilized by 
the $\varepsilon$-mechanism are admittedly cooler than the star. 
Although further 
exploration by considering stellar models with a range of H envelope masses,
stellar masses and effective temperatures, as well as higher degree
($\ell=3,4$) modes are needed to be more conclusive, this agreement
constitutes the first evidence of the presence of low-order
$g$-modes powered by the $\varepsilon$-mechanism in low-mass WDs,
thus giving strong observational support to the existence of stable
H burning in low-mass WDs.

\acknowledgments

Part of this work was
supported by AGENCIA through the Programa de Modernizaci\'on Tecnol\'ogica
BID 1728/OC-AR, and by the PIP 112-200801-00940 grant from CONICET.
This research has made use of NASA's Astrophysics Data System.

\bibliography{ms}

\end{document}